# S-matrices from 4d worldvolume

Warren Siegel

*CNYITP*
*Stony Brook University*

December 23, 2020

**Abstract**

We give an example of how conformal field theory methods in worldvolumes of dimension $d > 2$ could be used to calculate string-like amplitudes. The worldvolume propagator's logarithmic behavior is based on the use of worldvolume superspace (rather than the worldvolume ghost coordinates of a previous paper). Massless states have maximum spin $d$. Unitarity is not studied.

We also touch on some related topics for ordinary spinning strings: calculating closed-string trees in worldsheet superspace, and zeroth-quantization of OSp(1|2) with zero-mode ghosts only.

# Contents

# 1 Zeroth-quantized ghosts

In a previous paper [1] we considered the use of ghosts of zeroth-quantization to effectively reduce the dimension $d$ of worldvolumes to 2 by the Parisi-Sourlas method [2]. The $d-2$ fermionic worldvolume coordinates led to worldvolume propagators logarithmic in supercoordinates

$$\langle XX \rangle \sim \int d^{d|d-2}p\, e^{ip\cdot\sigma} \frac{1}{p^2} = -\ln(\sigma^2)$$

(e.g., using Schwinger parametrization, see below, and dimensional regularization, with appropriately normalized measure), where $\sigma^2$ is OSp($d|d-2$) covariant. (We use lower-case "$d$" for the worldvolume dimension to distinguish it from the spacetime dimension "$D$", even though the latter doesn't appear in this paper, for emphasis.) Tachyon amplitude calculations go as described there for (3 and) 4-point, reproducing the usual string results.

The same methods show the same is true for (closed string) $N$-point functions

$$A_N = \sigma_{1,N-1}^2 \sigma_{N-1,N}^2 \sigma_{N,1}^2 \int d^{(N-3)(d|d-2)}\sigma \prod_{i<j} (\sigma_{ij}^2)^{-\alpha_{ij}}$$

where $\sigma_{ij} \equiv \sigma_i - \sigma_j$. The measure factors are easily produced by exponentials of bosonic ghost fields, but like this $X$ they don't separate into "left" and "right", which are undefined: The exponentiated ghosts don't give fermionic fields, but only the analog of $c\bar{c}$ of the usual string.

We also use the following (collected here for later use):

$$\kappa \equiv \begin{cases} 1 & (open) \\ \tfrac{1}{2} & (closed) \end{cases}, \quad \alpha_{ij} \equiv -2\kappa^2 \alpha' k_i \cdot k_j$$

$$\langle X(i)\,X(j)\rangle = -\kappa\alpha' \ln|z_{ij} + ...|^2, \quad \alpha(s) = \kappa\alpha' s + 1/\kappa', \quad m_{tach}^2 = \begin{cases} -1/\kappa^2\alpha' & (bos.) \\ -1/2\kappa^2\alpha' & (RNS) \end{cases}$$

We exponentiate the worldvolume propagators $(\sigma_{ij}^2)^{-\alpha_{ij}}$ with Schwinger parameters (as used in deriving the Virasoro amplitude originally [3])

$$(\sigma^2)^{-\alpha} = \frac{1}{\Gamma(\alpha)} \int_0^\infty d\tau\, \tau^{\alpha-1} e^{-\tau\sigma^2}$$

As usual, the number of such factors can be reduced first by using conformal invariance to fix points $1, N-1, N$ (see below). All $d$-dependence for these amplitudes is contained in only the $\sigma$'s, which now appear only in Gaussian integrals, but after completing the square (and diagonalization) these all reduce to the usual

$$\int d^{d|d-2}\sigma\, e^{-a\sigma^2/2} = \frac{1}{a}$$

because of cancelation of $d-2$ bosonic integrals by fermionic.

The goal is that for excited states, where the OSp indices would appear (especially the SO parts) in vertex operators such as $\partial X$, more symmetry would be manifest than in the usual string formalism, such as nonperturbative dualities. (In particular, in F-theory worldvolume and spacetime indices are directly related [4].)

## 2 RNS string

### 2.1 Coordinates

In this paper we use a similar method, with the place of fermionic ghosts taken by the fermionic supercoordinates of worldvolume supersymmetry, as an analog to the Ramond-Neveu-Schwarz spinning string. (Superspace methods were applied to RNS amplitudes [5] before the advent of superspace in 4 spacetime dimensions. They were then derived from superspace propagators and vertex operators [6].)

We first "review" the worldsheet of the usual RNS string as superspace in terms of its realization of OSp(1|2) (over **R** for the open string boundary and **C** for the bulk/closed) as a projective space under (super) Möbius transformations.

We begin with

$$\zeta' = \mathcal{O}\zeta g: \quad \begin{pmatrix}\zeta_+ \\ \zeta_-\end{pmatrix}' = \begin{pmatrix}a & b \\ c & d\end{pmatrix}\begin{pmatrix}\zeta_+ \\ \zeta_-\end{pmatrix}g, \quad Z = \zeta_+\zeta_-^{-1} = \begin{pmatrix}z \\ \theta\end{pmatrix}$$

$$\Rightarrow \quad Z' = \frac{aZ+b}{cZ+d}, \quad \zeta'_- = (cZ+d)\zeta_-$$

where $\zeta$ is a (2|1) column vector, with global OSp(1|2) $\mathcal{O}$ (defining representation) acting linearly on the left, and local scale $g(\zeta)$ on "the right" (vacuous here, but relevant later), so $Z$ is (1|1) and local scale invariant. (Thus $a$ is (1|1) × (1|1), $b$ is (1|1) × 1, etc.) The global symmetry is defined by the invariance of the (hermitian, graded symmetric) OSp metric $M$: Introducing indices to keep grading signs straight,

$$\zeta'_M = \mathcal{O}_M{}^N \zeta_N$$

$$M^{MN} = (-1)^{MN} M^{NM}, \quad M = \begin{pmatrix}0 & 0 & -i \\ 0 & 1 & 0 \\ i & 0 & 0\end{pmatrix}$$

$$(\mathcal{O}^T M \mathcal{O})^{MN} \equiv (-1)^{P+MP} \mathcal{O}_P{}^M M^{PQ} \mathcal{O}_Q{}^N = M^{MN}$$

The $M$-index takes values $(\oplus, 0, \ominus)$ (which we'll avoid), with only 0 bosonic. (As usual, there is a hidden fermionic little-group index on the "twistor" $\zeta$.) We sandwiched the fermion $\theta$ between the bosons to order by conformal weight.

These transformations include "inversion"

$$\mathcal{O} = \begin{pmatrix} 0 & 0 & -1 \\ 0 & 1 & 0 \\ 1 & 0 & 0 \end{pmatrix} \quad \Rightarrow \quad z' = -\frac{1}{z}, \quad \theta' = \frac{\theta}{z}$$

("Inversion" in 2d, unlike higher d, refers to a continuous transformation: If the real axis of the complex plane is interpreted as the equator of the sphere, this is rotation about its axis by $\pi$. Thus, squaring it changes $\theta$ by a sign, as usual for spinors.) In terms of higher-dimensional superconformal identifications, the components of the OSp generators are

$$\begin{pmatrix} \Delta & Q & P \\ S & 0 & Q \\ K & S & \Delta \end{pmatrix}$$

($J$ is the imaginary part of $\Delta$; "0" is $R$-symmetry for 2d $\mathcal{N} = 1$.)

## 2.2 Densities

The basic idea is to first construct global invariants of OSp(1|2); then constructing local scale invariants from them is easy. Alternatively, fix the scale gauge $\zeta_- = 1$ for the global invariants, making them scalar densities of OSp. After applying momentum conservation (which implicitly invokes conformal weight as external particle mass), it's then manifest that the OSp transformations of the propagators cancel those of the vertex-operator $dz\, d\theta$ integrations. Invariance of $d\theta$ integrals without $dz$ is less clear, and is better treated by BRST (see below).

The OSp invariants are (in terms of "sigma teliko" $\varsigma$)

$$\zeta_i^T M \zeta_j = i\varsigma_{ij} \zeta_{i-} \zeta_{j-}, \quad \varsigma_{ij} = z_{ij} + i\theta_i \theta_j = -z_{ji}$$

$$\Rightarrow \quad \varsigma'_{ij} = \frac{\varsigma_{ij}}{(cZ_i + d)(cZ_j + d)}$$

(For inversion, $cZ + d = z$.) Similarly, the integration measure transforms as

$$d^2 Z' = \frac{d^2 Z}{cZ + d}$$

as a consequence of factoring out $d\zeta_-$. (The $\zeta_-$'s cancel between the boson and fermion in rescaling $d^2 \zeta_+ \to d^2 Z$.)

Tree amplitudes also have a single lone $d\theta$, for which it's simpler to look at just inversions. Then

$$d\theta = \frac{\partial}{\partial \theta} = \frac{\partial \theta'}{\partial \theta}\frac{\partial}{\partial \theta'} + \frac{\partial z'}{\partial \theta}\frac{\partial}{\partial z'} = \frac{d\theta'}{z}$$

$$\Rightarrow \quad d\theta' = zd\theta$$

Unfortunately, $d\theta$ in the absence of $dz$ isn't invariant under supersymmetry

$$q = \frac{\partial}{\partial \theta} + i\theta\frac{\partial}{\partial z}$$

However, the covariant spinor derivative

$$d = \frac{\partial}{\partial \theta} - i\theta\frac{\partial}{\partial z}$$

is. Furthermore, by the method just used both are covariant under inversions:

$$q' = zq, \quad d' = zd$$

Since $d$ is also supersymmetry and translation invariant, it follows by inversion (dilatation is trivial) that it must be a density (hence the term "covariant") under full OSp:

$$d' = (cZ + d)d$$

## 2.3 Projective BRST

The OSp(1|2) invariance of the vacuum (which will generalize to superconformal in higher worldvolume dimensions) allows fixing the positions of 3 $z$'s and 2 $\theta$'s of the vertices. This gauge (coordinate) fixing can be performed by a simplified version of Becchi-Rouet-Stora-Tyutin methods, applied to OSp only, and thus introducing ghosts for just that symmetry (i.e., the zero-modes of the usual ghosts). As a result, it can be expressed as "zeroth-quantization", expressing the transformations directly on the worldsheet coordinates instead of the fields. Thus effectively, instead of using gauge-(OSp(1|2)-)invariant, integrated vertex operators and BRST-invariant unintegrated ones, we use only integrated vertex operators, but with BRST-invariant insertions that kill some integrals. This allows a bit more flexibility in gauge fixing, as we'll see below.

Gauge fixing takes the standard form (in the absence of gauge averaging)

$$\delta(f)\delta([Q, f\})$$

inserted into the path integral, for gauge fixing function $f$. ($\delta$-functions follow from the ghost action after integrating out the antighost and Nakanishi-Lautrup nonminimal fields. Picture-changing insertions also take this form.) In this case, the $f$'s are 3 $z$'s and 2 $\theta$'s, minus the values where they're fixed, and the $[Q, f\}$'s are their infinitesimal OSp transformations, with parameters replaced by ghosts. Thus, integrating over the former $\delta$'s replaces those coordinates with their fixed values in the latter, and then integrating over the ghosts in the latter gives the measure as the Jacobian of the transformation from the fixed coordinates to the symmetry parameters [7] (i.e., the finite-dimensional Faddeev-Popov determinant).

Solving the linearized version of the graded orthogonality condition and substituting ghosts for parameters (in notation consistent with the usual mode expansion of ghost fields), the 0th-q BRST operator is

$$Q = c_1 \partial_z + c_0(z\partial_z + \tfrac{1}{2}\theta\partial_\theta) + c_{-1}z(z\partial_z + \theta\partial_\theta)$$
$$+\gamma_{1/2}(\partial_\theta + i\theta\partial_z) + \gamma_{-1/2}z(\partial_\theta + i\theta\partial_z) + \text{pure ghost terms}$$

with ghosts multiplying $P, \Delta, K, Q, S$, in that order. These 0th-q ghosts should be considered worldvolume ghost coordinates, partners to $z$ and $\theta$: S-matrices have (non-path) integrals over supercoordinates $dz_i\, d\theta_i$ for each vertex, and over superghosts $d^3c\, d^2\gamma$ overall.

From the 2 factors of

$$\delta(\theta)\delta(\{Q, \theta\}) = \theta\delta(\gamma_{1/2} + \gamma_{-1/2}z)$$

we then get

$$\frac{\theta_i\theta_j\delta(\gamma_{1/2})\delta(\gamma_{-1/2})}{z_{ij}}$$

These reduce the $z$ BRST transformation to the bosonic result

$$[Q, z] = c_1 + c_0 z + c_{-1}z^2$$

yielding the usual $z = \mathring{z}$ fixing

$$\delta(z_i - \mathring{z}_i)\delta(z_j - \mathring{z}_j)\delta(z_k - \mathring{z}_k)c_1 c_0 c_{-1}\mathring{z}_{ij}\mathring{z}_{jk}\mathring{z}_{ki}$$

Note the similarity of $\{Q, \theta\}$ and $[Q, z]$ to the zero-mode terms in the usual ghost fields.

BRST allows gauges that are more general, and sometimes more convenient, than those whose superconformal invariance is more manifest (see below).

## 2.4 Open

We now describe some amplitude calculations. Although we'll focus on the closed string, we warm up with open-string tachyon amplitudes. The tachyon vertex is the usual $e^{ik\cdot X}$, but $X(z, \theta)$ is now a superfield, with propagator

$$\Delta_{ij} \equiv \langle X(i)\, X(j)\rangle = -2\alpha' \ln \varsigma_{ij}$$

We fix OSp(1|2) invariance by choosing

$$z_1 = 0, \ z_{N-1} = 1, \ z_N \to \infty; \quad \theta_{N-1} = \theta_N = 0$$

Tachyon amplitudes are then

$$A_N = \varsigma_{1,N-1}\varsigma_{N,1} \int d^{N-3}z \, d^{N-3}\theta \, d\theta_1 \prod_{i<j} (\varsigma_{ij})^{-\alpha_{ij}}$$

(The $\varsigma$'s in the measure are purely bosonic due to the death of 1 of the $\theta$'s in each factor.) OSp(1|2) invariance is satisfied, for $m^2 = -1/2\alpha'$ (1/2 the bosonic string's tachyon). Inversion invariance is easily checked: As usual, the propagators generate $\prod z$ after applying momentum conservation (not squared as in the bosonic string, because this tachyon has ½ the squared mass), $N-3$ of which are canceled by $dZ$'s, the remaining by

$$\varsigma_{1,N-1}\varsigma_{N,1} d\theta_1 \to \frac{1}{z_1 z_{N-1}} \frac{1}{z_N z_1} z_1 = \frac{1}{z_1 z_{N-1} z_N}$$

For the 4-tachyon amplitude we then have

$$A_4 = \int dz \, d\theta_2 \, d\theta_1 \, (z + i\theta_1\theta_2)^{-\alpha(s)}(1-z)^{-\alpha(t)}$$

where

$$\alpha(s) = \alpha' s + 1$$

is the open-string leading trajectory. (This is the same trajectory in the NS string as the bosonic string, as both have a massless vector, but for NS the "$\mathfrak{F}_1$" tachyon on this trajectory is missing, and the usual $\mathfrak{F}_2$ tachyon appears on the odd-G-parity, first-daughter trajectory. Since these odd-G-parity tachyons appear on all the external lines in this amplitude, only even-G-parity states appear internally, by G-parity conservation.)

The $\theta$ integrals are trivial, and the $z$ integral yields a Beta function:

$$A_4 = -i\alpha(s) \frac{\Gamma[-\alpha(s)]\Gamma[1-\alpha(t)]}{\Gamma[1-\alpha(s)-\alpha(t)]} = i \frac{\Gamma[1-\alpha(s)]\Gamma[1-\alpha(t)]}{\Gamma[1-\alpha(s)-\alpha(t)]}$$

As an alternate gauge we consider

$$z_1 = 0, \ z_3 = 1, \ z_4 \to \infty; \quad \theta_2 = \theta_4 = 0$$

($\theta_2 = 0$ instead of $\theta_3$.) Then the amplitude is manifestly $s \leftrightarrow t$ symmetric, and the $d\theta$ integral factorizes from the $dz$:

$$A_4 = \int dz \, z^{-\alpha(s)}(1-z)^{-\alpha(t)} \int d\theta_3 \, d\theta_1 \, (1 + i\theta_1\theta_3)^{-\alpha(u)}$$

There is initially a change in measure

$$\varsigma_{13}\varsigma_{41} \to \frac{\varsigma_{13}\varsigma_{34}\varsigma_{41}}{\varsigma_{24}}$$

that becomes the same as $z_4 \to \infty$. The result for the amplitude is the same:

$$A_4 = -i\alpha(u)\frac{\Gamma[1-\alpha(s)]\Gamma[1-\alpha(t)]}{\Gamma[2-\alpha(s)-\alpha(t)]} = i\frac{\Gamma[1-\alpha(s)]\Gamma[1-\alpha(t)]}{\Gamma[1-\alpha(s)-\alpha(t)]}$$

using

$$\alpha(s) + \alpha(t) + \alpha(u) = 1$$

## 2.5 Closed

In the closed-string case, skipping directly to the 4 closed-string-tachyon amplitude, we have

$$A_4 = \int d^2z\, d^2\theta_2\, d^2\theta_1\, (|z + i\theta_1\theta_2|^2)^{-\alpha(s)/2}(|1-z|^2)^{-\alpha(t)/2}$$

where $m^2 = -2/\alpha'$, and

$$\alpha(s) = \tfrac{1}{2}\alpha's + 2$$

(We could choose a complex gauge like $\theta_3 = \bar\theta_1 = 0$, with the same result.)

The form of the amplitude resembles a 1-loop graph with $z$ as the loop momentum, and 2d vectors added to $z$ as the external momenta, but strange powers of $z$ for propagators. (There is also a less familiar dual interpretation of a position-space vertex diagram with external lines whose endpoints are given by the 2d vectors, and $z$ as the vertex.) This is essentially a propagator-correction graph.

The steps in evaluating this amplitude are therefore, in order:

1. Schwinger parametrize each amplitude factor (as, e.g., for the 0th-q OSp model above),
2. $\int dz$ (again Gaussian),
3. $\int d\theta$ (essentially a finite Taylor expansion),
4. $\int d\lambda$ (the scaling parameter),
5. $\int d\alpha$ (Feynman parameter).

By "momentum" conservation, the result of $z$ integration is in terms of the external momentum

$$(z + i\theta_1\theta_2) + (1-z) = 1 + i\theta_1\theta_2$$

Expanding the exponent $-\lambda\alpha(1-\alpha)|1+i\theta_1\theta_2|^2$ in $\theta$'s yields 2 terms at order $\theta^4$: Performing all integrals,

$$A_4 = \frac{B[1-\tfrac{1}{2}\alpha(s), 1-\tfrac{1}{2}\alpha(t)]}{\Gamma[\tfrac{1}{2}\alpha(s)]\Gamma[\tfrac{1}{2}\alpha(t)]} \{\Gamma[\tfrac{1}{2}\alpha(s)+\tfrac{1}{2}\alpha(t)] - \Gamma[1+\tfrac{1}{2}\alpha(s)+\tfrac{1}{2}\alpha(t)]\}$$

Using

$$\alpha(s) + \alpha(t) + \alpha(u) = 2$$

we have

$$A_4 = \frac{\Gamma[1-\tfrac{1}{2}\alpha(s)]\Gamma[1-\tfrac{1}{2}\alpha(t)]\Gamma[1-\tfrac{1}{2}\alpha(u)]}{\Gamma[1-\tfrac{1}{2}\alpha(s)-\tfrac{1}{2}\alpha(t)]\Gamma[1-\tfrac{1}{2}\alpha(s)-\tfrac{1}{2}\alpha(u)]\Gamma[1-\tfrac{1}{2}\alpha(t)-\tfrac{1}{2}\alpha(u)]}$$

Since the $\theta$ and $\bar{\theta}$ integrals already separate, we could also apply the method of Kawai, Lewellen, and Tye to factorize the amplitude into open-string amplitudes. (However, the same is not true for general $d$.)

The alternative gauge used for the open string produces a simpler calculation, again factorizing into a bosonic-string calculation times the $d\theta$ integrals:

$$A_4 = \int d^2z \, (|z|^2)^{-\alpha(s)/2}(|1-z|^2)^{-\alpha(t)/2} \int d\theta_3 \, d\theta_1 \, (1+i\theta_1\theta_3)^{-\alpha(u)/2} \int d\bar\theta_3 \, d\bar\theta_1 \, (1+i\bar\theta_1\bar\theta_3)^{-\alpha(u)/2}$$

The result then has a single term

$$A_4 = \frac{B[1-\tfrac{1}{2}\alpha(s), 1-\tfrac{1}{2}\alpha(t)]}{\Gamma[\tfrac{1}{2}\alpha(s)]\Gamma[\tfrac{1}{2}\alpha(t)]} \Gamma[\tfrac{1}{2}\alpha(s)+\tfrac{1}{2}\alpha(t)-1][-\alpha(u)]^2$$

yielding the same final result.

## 3 Higher worldvolume dimensions

### 3.1 Propagators

To find the simplest nontrivial examples, we look for first-quantized fields with logarithmic propagators, to produce amplitude integrands similar to those in string theory. The simplest generalization to arbitrary $d$ is for the anticommuting coordinates to appear only as the argument of the log, as the Lorentz invariant square of the translation and supersymmetry invariant combination

$$\mathring{\varsigma}_{ij} \equiv \sigma_{ij} + \tfrac{1}{2} i\theta_i \gamma \theta_j$$

(An alternative would be for some to appear in anticommuting $\delta$ functions; this would reproduce the requirement $d=2$ if they all appeared that way.) Actually this choice can be generalized: In

$$d^n\delta^n(\theta_{ij})f(\sigma_{ij}) = f(\varsigma_{ij}), \quad \varsigma_{ij} = \mathring{\varsigma}_{ij} + \tfrac{1}{2} i\bar\theta_{ij} M\gamma \theta_{ij}$$

(where $d$'s are understood to act on $\sigma_i$, $\theta_i$), the explicit choice of matrix $M$ (for 4d, $\sim \gamma_5$; see below) depends on the ordering of supersymmetry covariant spinor derivatives $d$ (whose total number of components is $n$). Only the totally antisymmetric ordering gives the combination $\mathring{\varsigma}$ given above.

Since this requires $X$ to be dimensionless (conformal weight 0) in the propagator $\Delta = \langle X X \rangle = -\ln$, and since supersymmetric fermionic derivatives $d$ (and integrals $d\theta$) have ½ the dimension of $\sigma$ derivatives, we look for gauge-fixed actions of the form

$$S \sim \int d^d\sigma\, d^d\theta\, X d^d X$$

where

$$(d^d)^2 \approx \Box^{d/2} \Rightarrow \Delta_{ij} \approx (d^d)^{-1}\delta^d(\theta_{ij})\delta^d(\sigma_{ij}) \approx (d^d)\Box^{-d/2}\delta^d(\theta_{ij})\delta^d(\sigma_{ij}) \approx \ln \varsigma_{ij}^2$$

I.e., the commuting and anticommuting dimensions are equal. Here "$\approx$" means up to gauge fixing.

We have neglected consideration of R-symmetry coordinates.

## 3.2 BRST

We look for superconformally invariant models. Then we can use this invariance to fix the positions of some of the vertex operators in scattering amplitudes: As in the case of 0th-q ghosts, we can use translations to fix 1 $\sigma$, conformal boosts to fix a second, and scale plus part of Lorentz to fix a third to be a unit vector "**1**" in a preferred direction. In addition, we can now use supersymmetry to kill 1 $\theta$, and S-supersymmetry to kill a second. All this is a direct generalization of the usual RNS string to higher dimensions.

But we can go further in $d > 2$: After fixing the third $\sigma$, we can use the residual $SO(d-1)$ in 4-and-higher-point amplitudes to rotate a fourth $\sigma$ to 2 dimensions (1 in the direction of **1**). This leaves a residual $SO(d-2)$ to apply to a fifth $\sigma$ in 5-point, etc.

In particular, this means that the 4-point amplitude has only a $d^2\sigma$ integration (besides the $d^{2d}\theta$), like the usual string. This differs from the case of 0th-q ghosts, which has effectively $d^{2(N-3)}\sigma$ for all $N$.

We will use the BRST method for generality. As for $d = 2$, gauge fixing the $\theta$'s first completely takes care of them and the $\gamma$'s with $\theta\delta(\{Q,\theta\})$, reducing the rest to the same as the bosonic case:

$$[Q_B, \sigma^a] = c_1^a + c_0\sigma^a + c_0^{[ab]}\sigma_b + c_{-1}^b(\sigma_b\sigma^a - \tfrac{1}{2}\delta_b^a\sigma^2)$$

(We again ignore R-symmetry.)

To simplify, we immediately fix the values of the $\sigma$'s, again with $\delta(\sigma - \mathring{\sigma})[Q, \sigma]$: The steps are then

1. Fix $\theta_2 = \theta_N = 0 \Rightarrow \gamma_{1/2} = \gamma_{-1/2} = 0$ (supersymmetry and S-supersymmetry). In the limit below, this gives the measure factor $[(\sigma_N)^2]^{-d/2}$.

2. Fix $\sigma_1 = 0 \Rightarrow c_1 = 0$ (translations).

3. Fix $\sigma_N = \infty \Rightarrow c_{-1} = 0$ (conformal boosts). This produces a measure factor of $[(\sigma_N)^2]^d$ (in the limit).

4. Fix $\sigma^a_{N-1} = \delta^a_0 \Rightarrow c^{[0a]}_0 = c_0 = 0$ (Lorentz boosts and dilatations).

5. Fix $\sigma^a_2 = (\Re z, \Im z, 0, ..., 0) \Rightarrow c^{[1a]}_0 = 0$ (SO($d-1$)/SO($d-2$) part of rotations). This gives the measure factor $(\Im z)^{d-2}$.

6. Etc. (for $N > 4$).

(Naming the 2 surviving components of $\sigma_2$ "$z$" is notation, to make things similar to $d = 2$.) For $d = 3$ (simple superconformal symmetry OSp(1|4)), this would fix all superconformal symmetry; for $d > 3$ we're left with SO($d+2-N$) and R-symmetry, which we'll ignore (at least for the 4-pt. function).

As for the case of 0th-q ghosts, the measure factors can be produced by exponentials of ghost bosons with propagator similar to that of $X$, one for $(\varsigma^2)^{d/2}$'s between 1, $N-1$, and $N$ (associated with fixing $\sigma$'s), and one with opposite signature to cancel that between $N-1$ and $N$ (associated with fixing $\theta$'s). Also, because they generate only $(\varsigma^2)^{d/2}$ (or the inverse), the exponentials are themselves bosons. (They correspond to (ghost)$^d$ for $d$ even, generalizing from 2d left ghost × right.)

## 3.3  4-point

For tachyons, we again use $e^{ik \cdot X}$ as vertex operator. We have tacked on a spacetime vector index to $X$, as in ordinary string theory. (Especially in the vector multiplet case, this can be interpreted as a gauge group index.) This may be peculiar if $X$ is a gauge field; in F-theory it leads to Gauss-law constraints on the external wave functions [4].

Using the simpler gauge ($\theta_2 = 0$ and not $\theta_3$), the generalization of the 4-tachyon closed-string amplitude should then look like

$$A_4 = \int d^2 z\, (|z|^2)^{-\alpha(s)/2} (|1-z|^2)^{-\alpha(t)/2} (z - \bar z)^{d-2} \int d^d\theta_3\, d^d\theta_1\, [(\mathbf{1} + \tfrac{1}{2} i \theta_1 \gamma \theta_3)^2]^{-\alpha(u)/2}$$

where now

$$\alpha(s) = \tfrac{1}{2} \alpha' s + d, \quad m^2_{tach} = -\frac{d}{\alpha'}$$

The change in $m^2$ is from the conformal weight of the vertex operator, due to $\int d^d\sigma\, d^d\theta$; the change in $\alpha_0$ then follows from rewriting $k_i \cdot k_j$ in terms of $m^2$ and $s_{ij}$. Thus the spins of the massless states run up to $d$.

Schwinger parametrizing, with

$$\tau_1 = \lambda\alpha, \ \tau_2 = \lambda(1-\alpha)$$

we have

$$A_4 = \frac{\int d^{2d}\theta \ldots}{\Gamma[\tfrac{1}{2}\alpha(s)]\Gamma[\tfrac{1}{2}\alpha(t)]} \int_0^1 d\alpha\, \alpha^{\alpha(s)/2-1}(1-\alpha)^{\alpha(t)/2-1}$$

$$\times \int_0^\infty d\lambda\, \lambda^{[\alpha(s)+\alpha(t)]/2-1} e^{-\lambda\alpha(1-\alpha)} \int d^2\sigma\, \sigma_1^{d-2} e^{-\lambda(\sigma_0^2+\sigma_1^2)}$$

Note that

$$\int d^2\sigma\, \sigma_1^{d-2} e^{-\lambda(\sigma_0^2+\sigma_1^2)} \sim \lambda^{-d/2} \sim \int d^d\sigma\, e^{-\lambda\sigma^2}$$

so that gauge fixing SO($d-1$) was unnecessary. Using this result and rescaling $\lambda$'s to factorize its integral from the Feynman parameter's,

$$\hat{\lambda} = \lambda\alpha(1-\alpha)$$

and applying

$$\alpha(s) + \alpha(t) + \alpha(u) = d$$

we have

$$A_4 = \frac{\int d^{2d}\theta \ldots}{\Gamma[\tfrac{1}{2}\alpha(s)]\Gamma[\tfrac{1}{2}\alpha(t)]} \int_0^1 d\alpha\, \alpha^{[d-\alpha(t)]/2-1}(1-\alpha)^{[d-\alpha(s)]/2-1} \int_0^\infty d\hat{\lambda}\, \hat{\lambda}^{-\alpha(u)/2-1} e^{-\hat{\lambda}}$$

From Appendix A, we find

$$\int d^d\theta_3\, d^d\theta_1\, [(1+\tfrac{1}{2}i\theta_1\gamma\theta_3)^2]^{-\alpha(u)/2} \sim \frac{\alpha(u)\Gamma\left[\tfrac{d}{2} - \tfrac{1}{2}\alpha(u)\right]}{\Gamma\left[1 - \tfrac{d}{2} - \tfrac{1}{2}\alpha(u)\right]}$$

Putting everything together,

$$A_4 = \prod_{r=s,t,u} \frac{\Gamma\left[\tfrac{d}{2} - \tfrac{1}{2}\alpha(r)\right]}{\Gamma[\tfrac{1}{2}\alpha(r)]} = \prod_{r=s,t,u} \frac{\Gamma\left[-\tfrac{1}{4}\alpha' r\right]}{\Gamma\left[\tfrac{d}{2} + \tfrac{1}{4}\alpha' r\right]}$$

It's then easy to check (e.g., using the Sterling approximation) that as $s \to \infty$, $A_4 \sim s^{\alpha(t)}$ as usual. (This confirms the identification of $\alpha(t)$ as spin. Also, use of $\sum \alpha = d$ to obtain an $stu$-symmetric result confirms the choice of $\alpha_0$.)

The spectrum of internal states on the leading trajectory is the usual

$$\alpha' m^2 = 0, 4, 8, \ldots$$

This implies a generalization of G-parity. However, $\alpha(m^2) = 0$ ("$\mathfrak{F}_1$ vacuum") at $\alpha' m^2 = -2d$, and the tachyon is at $-d$. This suggests a generalization of the Virasoro constraint that worldsheet left and right modes have equal levels: equal mode levels for each of the $d$ directions. The corresponding worldvolume metric whose zero-modes yield these constraints would need only $d$ components, 1 for gauging transformations of each worldvolume coordinate. (The time component zero-mode would give the Klein-Gordon equation.) Such a situation exists in F-theory [4].

# 4  4d worldvolume

## 4.1  Coordinates

Because the fermionic derivative is always a worldvolume-Lorentz spinor, the fermionic dimension is always a power of 2, a power that rises exponentially with the bosonic dimension. In Minkowski space, the only case where both dimensions can be equal for $d > 2$ is $d = 4$. Furthermore, superconformal groups don't exist for $d > 6$ [8]. We'll restrict ourselves to $\mathfrak{N} = 1$ to avoid R-symmetry coordinates.

Chiral propagators are superconformal by the usual projective superspace constructions, that generalize to higher $\mathcal{N}$ [9]. (Earliest versions of superspace superconformal transformations in 4 dimensions [10, 11] didn't take advantage of their simplification in chiral superspaces [12].) We start with (cf. 2d above) a $(4|1) \times 2$ matrix with global SU(2,2|1) $U$ acting linearly on the left and local GL(2,C) $g(\zeta)$ on the right, so $Z$ is $(2|1) \times 2$ and GL(2,C) invariant (see [13] for notation):

$$\zeta_M{}^\beta = \begin{pmatrix} \zeta_{\dot\mu}{}^\beta \\ \zeta_0{}^\beta \\ \zeta_{-\mu}{}^\beta \end{pmatrix}$$

$$\zeta' = U\zeta g: \quad \begin{pmatrix} \zeta_+ \\ \zeta_- \end{pmatrix}' = \begin{pmatrix} a & b \\ c & d \end{pmatrix} \begin{pmatrix} \zeta_+ \\ \zeta_- \end{pmatrix} g, \quad Z = \zeta_+ \zeta_-^{-1} = \begin{pmatrix} z^\beta{}_{\dot\alpha} \\ \theta^\beta \end{pmatrix}$$

$$\Rightarrow \quad Z' = (aZ + b)(cZ + d)^{-1}, \quad \zeta'_- = (cZ + d)\zeta_-$$

$U$ is (pseudo)unitary with respect to the metric $\Upsilon$:

$$(\overline{U}\Upsilon U)^{\overline{M}N} \equiv (-1)^{\overline{P}} \overline{U}^{\overline{M}}{}_{\overline{P}} \Upsilon^{\overline{P}Q} U_Q{}^N = \Upsilon^{\overline{M}N}, \quad \Upsilon = \begin{pmatrix} 0 & 0 & iC^{\mu\nu} \\ 0 & 1 & 0 \\ -i\overline{C}^{\dot\mu\dot\nu} & 0 & 0 \end{pmatrix}$$

Similar remarks apply to inversion (from which the above can also be derived). If the conformal group is interpreted as the Lorentz group in 1 extra space + 1 extra time dimension, then inversion is time reversal with respect to the extra time. In terms of $\zeta$ (which is the supersymmetric extension of the spinor representation of this Lorentz), this means complex conjugation, combined with switching $\zeta_-$ with the bosonic part of $\zeta_+$ (both square matrices):

$$\zeta' = \mathcal{M}\zeta^* m \, ; \qquad \mathcal{M}_M{}^N = \begin{pmatrix} 0 & 0 & I \\ 0 & -i & 0 \\ I & 0 & 0 \end{pmatrix}, \qquad m_{\dot\alpha}{}^\beta = \delta_{\dot\alpha}^\beta$$

($m$ is the unit vector in the time direction.) The result is ($\bar z \equiv z^\dagger$)

$$z' = \frac{\bar z}{\frac{1}{2}\bar z^2}, \qquad \theta' = -i\bar\theta \not{z}' \qquad (\bar\theta' = -i\theta\bar{\not{z}}')$$

$\zeta$ satisfies the reality condition (in spinor notation for $\sigma^{\alpha\dot\beta}, \theta^\alpha, \bar\theta^{\dot\beta}$)

$$(\bar\zeta\Upsilon\zeta)^{\dot\alpha\beta} \equiv (-1)^{\overline{M}}\bar\zeta^{\dot\alpha}{}_{\overline{M}}\Upsilon^{\overline{M}N}\zeta_N{}^\beta = 0 \quad \Rightarrow \quad \bar z - z = i\theta\bar\theta \quad \Rightarrow \quad z = \sigma - \tfrac{1}{2}i\theta\bar\theta, \quad \sigma = \sigma^\dagger$$

(In the coset approach, this is part of the unitarity condition on the group element of which $\zeta$ is a part.) This is a major distinction from the 2d OSp case.

## 4.2 Densities

The chiral integration measure transforms as (from scaling by $\zeta_-$ and factoring out $d\zeta_-$)

$$dZ' = \frac{dZ}{[\det(cZ + d)]^3}$$

and under inversion

$$dZ' = \frac{d\bar Z}{(\tfrac{1}{2}\bar z^2)^3}$$

where $dZ = d^4z\, d^4\theta$, $d\bar Z = d^4\bar z\, d^4\bar\theta$. If we treat $z$ and $\bar z$ as independent, the full integration measure can be written as

$$dZ\, d\bar Z\, \delta^4(\bar\zeta\Upsilon\zeta) = \frac{d\Sigma}{(\det \zeta_- \det \bar\zeta_-)^2}, \qquad d\Sigma \equiv d^4\sigma\, d^4\theta$$

It then transforms as

$$d\Sigma' = \frac{d\Sigma}{\det(cZ + d) \det(\bar c\bar Z + \bar d)}$$

and under inversion

$$d\Sigma' = \frac{d\Sigma}{\tfrac{1}{2}z^2\, \tfrac{1}{2}\bar z^2}$$

By a calculation similar to 2d, we find for inversions

$$d\theta' = -id\bar\theta\not{z} \quad \Rightarrow \quad d^2\theta' = -d^2\bar\theta\tfrac{1}{2}\bar z^2 \quad \Rightarrow \quad d^4\theta' = d^4\theta\tfrac{1}{2}z^2\,\tfrac{1}{2}\bar z^2$$

The global invariants are

$$\bar{\zeta}_j \Upsilon \zeta_i = i\bar{\zeta}_{-j} \varsigma_{ij} \zeta_{-i}$$

where

$$\varsigma_{ij} = z_i - \bar{z}_j + i\theta_i \bar{\theta}_j = \bar{\varsigma}_{ji} = \mathring{\varsigma}_{ij} - \tfrac{1}{2} i\theta_{ij} \bar{\theta}_{ij}, \qquad \mathring{\varsigma}_{ij} = \sigma_{ij} + \tfrac{1}{2} i(\theta_i \bar{\theta}_j + \bar{\theta}_i \theta_j)$$

Under superconformal transformations

$$(\varsigma'_{ij})^2 = \frac{(\varsigma_{ij})^2}{\det(cZ_i + d)\det(\bar{c}\bar{Z}_j + \bar{d})}$$

and under inversions

$$(\varsigma'_{ij})^2 = \frac{(\bar{\varsigma}_{ij})^2}{\tfrac{1}{2}\bar{z}_i^2 \tfrac{1}{2} z_j^2}$$

It's interesting to note that the chiral and antichiral superspaces $Z$ and $\bar{Z}$ are the direct analogs of the left and right-handed 2d ones, except that $\varsigma_{ij}$ and $\bar{\varsigma}_{ij}$ now connect chiral to antichiral instead of being purely one or the other.

A useful identity follows from

$$\Theta \equiv -\tfrac{1}{2} i\theta\bar{\theta} \quad \Rightarrow \quad \Theta^a \Theta^b = \tfrac{1}{4}\eta^{ab}\Theta^2, \quad \Theta^a \Theta^b \Theta^c = 0; \quad \Theta^2 = \delta^4(\theta)$$

$$\Rightarrow \quad z^2 \bar{z}^2 = (\sigma^2 + \Theta^2)^2 - 4(\sigma \cdot \Theta)^2 = \sigma^2(\sigma^2 + \Theta^2) = (\sigma^2 + \tfrac{1}{2}\Theta^2)^2$$

An almost identical calculation yields

$$(\varsigma_{ij})^2 (\bar{\varsigma}_{ij})^2 = [(\mathring{\varsigma}_{ij})^2 + \tfrac{1}{2}\Theta_{ij}^2]^2; \qquad \Theta_{ij} \equiv -\tfrac{1}{2} i\theta_{ij}\bar{\theta}_{ij}, \quad \Theta_{ij}^2 = \delta^4(\theta_{ij})$$

## 4.3 Multiplets

There is the usual complication that gauge fixing breaks conformal invariance off shell. At least for 4d $\mathcal{N} = 1$ the modification is rather simple: The gauge covariant and gauge fixing terms (which switch roles for the vector and scalar multiplets) are separately conformally covariant, so the transformations of log terms are adequate, in the same way as for ordinary strings. Even if propagators are not conformal, invariance may be restored at the end for correlators of conformal operators.

An example of $d > 2$ logarithmic propagators is known: 4d $\mathcal{N} = 1$ (minimal-)super Yang-Mills (vector multiplet) in the supersymmetric anti-Fermi-Feynman gauge (exercise VIB5.4 in [13]). This is the gauge where the sign for the Fermi-Feynman gauge-fixing term is reversed. This replaces the kinetic operator $\Box$ with $d^4$ with exactly antisymmetric ordering. (The analog in a Clifford algebra is "$\gamma_5$".) The same can be done for the chiral scalar superfield (scalar multiplet), expressed as $d^2$ on an unconstrained superfield, but now the role of gauge-invariant and gauge-fixing terms is switched.

The free, massless actions for 4d $\mathcal{N} = 1$ matter multiplets are

$$S_{CM} = -\int d\Sigma\, \overline{\varphi}\varphi = -\int d\Sigma\, \overline{\psi} d^2 \overline{d}^2 \psi = \int d\Sigma\, \overline{\psi} \Pi^{(-)}(-\tfrac{1}{2}\Box)\psi, \qquad \varphi = \overline{d}^2\psi$$

$$S_{VM} = -\int d^4\sigma\, d^2\theta\, W^2 = -\int d\Sigma\, \tfrac{1}{2} V d^\alpha \overline{d}^2 d_\alpha V = -\int d\Sigma\, \tfrac{1}{2} V \Pi^{(0)}(-\tfrac{1}{2}\Box)V,$$

$$W^\alpha = \overline{d}^2 d_\alpha V$$

in terms of off-shell projection operators:

$$\Pi^{(+)} = \frac{-\overline{d}^2 d^2}{-\tfrac{1}{2}\Box}, \quad \Pi^{(-)} = \frac{-d^2 \overline{d}^2}{-\tfrac{1}{2}\Box}, \quad \Pi^{(0)} = \frac{d^\alpha \overline{d}^2 d_\alpha}{-\tfrac{1}{2}\Box} = 1 - \Pi^{(+)} - \Pi^{(-)}$$

General propagators can be expressed as

$$\Delta(i, j) = \frac{\mathfrak{N}}{-\tfrac{1}{2}\Box} \delta^4(\theta_{ij})\delta^4(\sigma_{ij})$$

As a simple example of superconformal invariance, the interacting chiral multiplet

$$S = -\int d\Sigma\, \overline{\varphi}\varphi + \left(\lambda \int dZ\, \varphi^3 + h.c.\right)$$

is invariant under superconformal transformations

$$\varphi'(Z') = \det(cZ + d)\varphi(Z)$$

and inversions

$$\varphi'(Z') = \tfrac{1}{2} \overline{z}^2 \overline{\varphi}(\overline{Z})$$

as seen from the transformations of $dZ$ and $d\Sigma$ in the previous subsection.

## 4.4 Gauges

In Landau gauge the numerator $\mathfrak{N}$ is just one of the $\Pi$'s:

$$\text{Landau:} \quad \mathfrak{N} = \Pi^{(+)},\, \Pi^{(-)},\, -\Pi^{(0)}$$

Unfortunately, these propagators aren't useful.

There are 2 choices which have particularly simple forms. The first is a combination of chiral and antichiral, corresponding to an $X$ that is a sum of chiral and antichiral parts (analogous to holomorphic + antiholomorphic for $d = 2$):

$$\mathfrak{N} = -\tfrac{1}{4}\alpha'(\Pi^{(+)} + \Pi^{(-)}) \quad \Rightarrow \quad \Delta^{(c)} = \tfrac{1}{4}\alpha'\frac{\{\overline{d}^2, d^2\}}{(\tfrac{1}{2}\Box)^2}\delta^4(\theta_{ij})\delta^4(\sigma_{ij}) = -\tfrac{1}{4}\alpha' \ln[(\varsigma_{ij})^2(\overline{\varsigma}_{ij})^2]$$

This chiral propagator has the superconformal transformation

$$\Delta'^{(c)}_{ij} = \Delta^{(c)}_{ij} + \tfrac{1}{4}\alpha' \ln|\det(cz_i + d)\det(cz_j + d)|^2$$

and for inversion

$$\Delta'^{(c)}_{ij} = \Delta^{(c)}_{ij} + \tfrac{1}{4}\alpha' \ln|(\tfrac{1}{2} z_i^2)(\tfrac{1}{2} z_j^2)|^2$$

This non-invariance is of the same form as in normal string theory, and reduces by momentum conservation to weight factors that cancel with those of the measure.

For the second, we note that using unity in place of a projection operator, as for the vector multiplet in super Fermi-Feynman gauge, gives the simple (but useless for our purposes) propagator

$$\mathfrak{N} = -1 \quad \Rightarrow \quad \Delta^{(I)} = -\frac{\delta^4(\theta_{ij})}{\tfrac{1}{2}\sigma_{ij}^2} = -\ln\left(1 + \frac{\delta^4(\theta_{ij})}{\tfrac{1}{2}\sigma_{ij}^2}\right)$$

We then note

$$(\varsigma_{ij})^2(\overline{\varsigma}_{ij})^2 = (\mathring{\varsigma}_{ij})^4\left(1 + \frac{\delta^4(\theta_{ij})}{\tfrac{1}{2}\sigma_{ij}^2}\right)^{1/2}$$

Thus in the anti-FF gauge,

$$\mathfrak{N} = -\tfrac{1}{8}\alpha'(-\Pi^{(0)} + \Pi^{(+)} + \Pi^{(-)}) = -\tfrac{1}{8}\alpha'[-1 + 2(\Pi^{(+)} + \Pi^{(-)})] \quad \Rightarrow \quad \Delta^{(0)} = -\tfrac{1}{2}\alpha' \ln \mathring{\varsigma}_{ij}^2$$

Unfortunately, both of these have the wrong sign to correspond to gauge fixing of the usual, unitary actions. So we consider more general covariant gauges: For positive $a$,

$$\text{SM:} \quad (\Pi^{(+)} + \Pi^{(-)}) + (a+1)\Pi^{(0)} = -a(\Pi^{(+)} + \Pi^{(-)}) + (a+1)$$
$$\text{VM:} \quad -\Pi^{(0)} - (a+1)(\Pi^{(+)} + \Pi^{(-)}) = -a(\Pi^{(+)} + \Pi^{(-)}) - 1$$

After scaling fields appropriately, this gives propagators of the form

$$\mathfrak{N} = -\tfrac{1}{4}\alpha'[(\Pi^{(+)} + \Pi^{(-)}) + (\beta - \tfrac{1}{2})] \quad \Rightarrow \quad \Delta = -\tfrac{1}{2}\alpha' \ln[\mathring{\varsigma}_{ij}^2 + \beta\delta^4(\theta_{ij})]$$

## 4.5 Four-point

Superconformal invariance would be manifest for the scalar multiplet propagator in Landau gauge. The residual part of superconformal symmetry is SO(d–1) (SO(3) here) and R-symmetry (U(1) in this case). Each boson can be expressed as the sum of "left" (chiral, $\varsigma^2$) and "right" (antichiral, $\bar\varsigma^2$) parts, and thus the exponential as the product, but again their individual matrix elements connect chiral to antichiral.

We'll consider both the scalar and vector multiplet in various gauges by using the more general propagator (see above)

$$\Delta_{ij} = -\tfrac{1}{2}\alpha' \ln[\mathring\xi^2_{ij} + \beta\delta^4(\theta_{ij})]$$

The manifestly superconformal case is $\beta = 1/2$; the case we considered in our $d > 2$ generalization is $\beta = 0$. For the 4-tachyon amplitude, in either gauge we considered, all $\theta$ dependence was isolated in 1 factor. We then note

$$[\delta^4(\theta_{ij})]^2 = 0 = \delta^4(\theta_{ij})\theta_i\gamma\theta_j$$

Thus $\delta^4(\theta_{ij})$ does not contribute to $\int d^4\theta_i\, d^4\theta_j$ (which needs $\theta^8$), so the amplitude is $\beta$ independent. It follows that it:

- agrees with our $d > 2$ generalization above,
- is manifestly superconformal,
- works for either multiplet, and
- would be gauge independent if not for the $\beta$ dependence of the normalization of the propagator.

## 5 Conclusions

We have not analyzed the generalized (super) Virasoro constraints. These are necessary for studying unitarity. Their ghosts should have zero-modes corresponding to the zeroth-quantized ones we introduced for the amplitude measure, coming from the superconformal subalgebra of the Virasoro (general worldvolume coordinate) algebra. In particular, we could check consistency of massless spins > 2.

The physical bosons appear in the middle of the $\theta$ expansion of the prepotentials, making $\theta$ integration a proper normalization without picture changing or insertions [14].

The massless vertex should be $(d_\alpha X)^d$, as a generalization of 2d. However, it isn't invariant under the gauge transformations of either multiplet: For the scalar multiplet, the field strength is $d^2 X$, while for the vector multiplet, it's $d^3 X$.

One could also see if some of our assumptions that lead to only $d = 2$ or $4$ might be weakened, while still preserving our proposed result for general $d$. In particular, R-symmetry coordinates might play a crucial role.

These results should be gauge independent. This suggests looking at other gauges (such as Wess-Zumino), where the usual power (instead of log) propagators appear, and thus at (worldvolume) nonsupersymmetric theories, and in other dimensions.

## Acknowledgements

This work is supported by NSF grant PHY-1915093.

## A  Appendix

Here we evaluate
$$\int d^d\theta_3 \, d^d\theta_1 \, [(1+V)^2]^a$$
where
$$V \equiv \tfrac{1}{2} i\theta_1 \gamma \theta_3, \quad a = -\tfrac{1}{2}\alpha(u)$$
Since by Lorentz and Bose symmetry $V^{a_1}...V^{a_d}$ must be proportional to $\theta_1^d \theta_3^d$ times the product of $d/2$ $\eta$'s totally symmetrized in all vector indices, this calculation is most easily performed analytically: We thus need to evaluate
$$[(1 + 2\partial_0 + \Box)^a (\sigma^2)^{d/2}]|_{\sigma=0}$$

From
$$\Box(\sigma^2)^n = 4n(n + \frac{d}{2} - 1)(\sigma^2)^{n-1}$$
we find
$$\Box^n (\sigma^2)^{d/2} \sim \frac{4^n}{(d-1-n)!(\frac{d}{2}-n)!}(\sigma^2)^{d/2-n}$$
ignoring coefficients dependent on only $d$. Since we need only the terms
$$(1 + 2\partial_0 + \Box)^a \to \sum_{n=0}^{d/2} \Box^n (2\partial_0)^{d-2n} \binom{a}{d-n}\binom{d-n}{n}$$
we then have
$$a\sum_{n=0}^{d/2} \binom{a-1}{a-d+n}\binom{d/2}{n}$$

using
$$\partial_0^{d-2n} \sigma_0^{d-2n} = (d-2n)!$$

Again evaluating analytically, we write this as
$$a \oint_0 \frac{dx}{2\pi i x} x^{1-a}(1+x)^{a-1}(1+x)^{d/2}$$

This leads to the final result
$$\int d^d\theta_3 \, d^d\theta_1 \, [(\mathbf{1} + \tfrac{1}{2} i\theta_1\gamma\theta_3)^2]^{-\alpha(u)/2} \sim \frac{\alpha(u)\Gamma\left[\frac{d}{2} - \tfrac{1}{2}\alpha(u)\right]}{\Gamma\left[1 - \frac{d}{2} - \tfrac{1}{2}\alpha(u)\right]}$$